\documentstyle[preprint,eqsecnum,aps]{revtex}
\begin{document}
\draft
\title{Resonant scattering on impurities in the Quantum Hall Effect}
\author{S. A. Gurvitz}
\address{Department of Nuclear Physics, Weizmann Institute of
         Science, Rehovot 76100, Israel\\
and TRIUMF, Vancouver, B.C., Canada V6T\ 2A3}
\maketitle
\begin{abstract}
We develop a new approach to carrier transport
between the edge states via resonant scattering on
impurities, which is applicable
both for short and long range impurities.
A detailed analysis of resonant scattering
on a single impurity is performed. The results are used for
study of the inter-edge transport by multiple
resonant hopping via different impurities' sites.
It is shown that the total conductance can be found from an
effective Schroedinger equation with constant
diagonal matrix elements in the Hamiltonian, where the complex
non-diagonal matrix elements are the
amplitudes of a carrier hopping between different impurities.
It is explicitly demonstrated
how the complex phase leads to Aharonov-Bohm oscillations in the total
conductance. Neglecting the contribution of self-crossing
resonant-percolation trajectories, one finds
that the inter-edge carrier transport is similar to propagation in
one-dimensional system with off-diagonal disorder. We demonstrated
that each Landau band has an extended state
$\bar E_N$, while all other states are localized. The localization
length behaves as $L_N^{-1}(E)\sim (E-\bar E_N)^2$.
\end{abstract}

\section{Introduction}

The quantum Hall effect (QHE) is characterized by plateaus of
zero longitudinal resistance and quantized Hall resistance as a
function of the magnetic field. When the Fermi level passes
through the plateaus, the QHE breaks down, and the
longitudinal resistance appears. The explanation of the integer QHE in a
the framework of the Landauer approach\cite{str} is based on the suppression of
backscattering in a strong magnetic field\cite{but}: a carrier
moving along the boundary of a sample cannot reverse direction of its motion
if scattered by an impurity, unless it jumps to
another boundary. If a sample is much wider then the magnetic length, the
inter-edge scattering is likely to be negligible and the conductance is
expected to be quantized. However, when the Fermi energy is near Landau level
centers, the electron states extend across the sample. As a result, the
inter-edge current appears and so does the longitudinal resistance.

The localization and scaling properties of a disordered 2D electron gas
near the Landau level in the presence of a
strong external magnetic field has been a subject of many numerical
\cite{liu} and analytical studies\cite{avi}. Although various computer
simulations strongly support the concept of the power-law divergence of the
localization length near the Landau level
center\cite{liu}, the question as to the nature of the delocalization is not
yet resolved.

Actually, the strong-field Landau localization is related to the
2D scattering problem on randomly distributed impurities.
The problem can be essentially simplified if the electron-impurity
scattering goes through quasi-stationary states near impurity sites.
In fact, this always happens when the electron energy passes the Landau
levels, since the impurities
generate quasi-stationary states near the Landau levels thresholds.
In this case one can retain only the resonance part
in the electron-impurity scattering amplitude,
instead of solving the 2D scattering problem in its full complexity.

We concentrate in this paper on the resonant scattering on impurities
in QHE in general, and in connection with the
localization properties of electronic
states near Landau levels. For treatment of this problem
we propose a new approach, which is
based on couple-channel multiple-scattering theory and
allows an analysis of QHE both in narrow and in wide samples. This approach
reduces 2D scattering problem to an effectively 1D problem,
and therefore makes it possible to investigate the localization and scaling
properties of a distorted 2D system analytically.

We start with a detailed investigation of the inter-edge transitions via
the resonance scattering on a single impurity. Although this process has
been discussed in the literature\cite{pran,jain,pokr,suk},
the analysis was always concentrated on a specific type of the
impurity potentials. Here we present an unified approach and derive simple
analytical expressions for the resonance energy and the partial widths.
However, the probability of such direct enter-edge
transitions drops down very rapidly with a sample width. Hence, the
resonant tunneling on a single impurity
can produce an appreciable effect only in narrow samples.
One can expect that in a
the case of wide samples the inter-edge
resonant transport goes via by multiple resonant
transitions (hopping) of a carrier via different impurities. In fact, this
process had not attracted an appropriate attention, except for a similar
problem of tunneling transparency of disordered systems (with no
magnetic field), which has been studied by Lifshitz and
Kirpichenkov\cite{kirp}.
In the second part of the paper we concentrate on
2D electron gas transport via subsequent resonant scattering on different
impurities in the presence of a strong magnetic field.
The electron-electron interaction is not considered here.

The paper is organized as follows: General description of the couple-channel
approach is given in Section 2. In Section 3 we present
a detailed analysis
of resonant scattering on a single impurity. The multiple
resonance scattering on two impurities is discussed in Section 4.
We singled out
this particular case in order to exemplify how our approach is
extended to any number of impurities participating in the resonant transport.
We also obtain there an analytical expression for the complex (off-diagonal)
amplitude describing the resonant hopping between different impurity
sites and for the diagonal energy shift. The resonant transport
via $n$ impurities is discussed in Section 5.
Section 6 is the summary. In the last Section we compare
our results for localization of the 2D
disordered electron gas in a strong magnetic field
with those obtained by different methods.

\section{Formulation of the problem}

Consider two-dimensional noninteracting electrons gas
in the presence of a perpendicular magnetic field {\mbox{\boldmath $B$}}.
The electrons are
confined along the $y$-direction and free to move
along the $x$-direction.
Consider also impurities inside the system.
The Schroedinger equation describing the electron motion in the
($x,y$)-plane for the Landau gauge,
{\mbox{\boldmath $A$}}=($-By,0,0$), is
\begin{equation}
\left\lbrack \frac{1}{2m}\left (
-i\hbar{d\over dx}+{e\over c}A_x\right )^2
-\frac{\hbar^2}{2m}{d^2\over dy^2}+W(y)
+V(x,y)\right\rbrack\Psi (x,y)=E\Psi (x,y)
\label{a1}
\end{equation}
where $m$ is the electron effective mass.
The potential $W(y)$ provides confinement of the electron motion
along the $y$-direction. For the definiteness we take it in a form
of confining walls, $W(y)=0$ for $|y|\leq a\;$ and
$\; W(y)=+\infty$ for $|y|>a$. The impurities inside the channel are
described by the potential $V(x,y)\equiv \sum_j V_j(x-x_j,y-y_j)$,
where $(x_j,y_j)$ are the coordinates of the impurity centers.

In the absence of impurities $(V=0)$ the solution of
Eq.\ (\ref{a1}) can be written as
\begin{equation}
\phi_{nk}(x,y)=e^{ikx}\Phi_{nk}(y)
\label{a2}
\end{equation}
where $\Phi_{nk}$ satisfies the reduced equation
\begin{equation}
\left\lbrack -\frac{\hbar^2}{2m}{d^2\over dy^2}+{m\omega_c^2(y-\ell^2k)^2
\over 2}+W(y)\right\rbrack\Phi_{nk}(y)=E_{nk}\Phi_{nk}(y)
\label{a3}
\end{equation}
with $\omega_c=eB/mc$ is the cyclotron frequency,
$\ell\ =\sqrt{\hbar c/eB}$ is the magnetic length and
\begin{equation}
E_{nk}=E_n+{\cal K}_n(k)
\label{a4}
\end{equation}
The energy spectrum $E_{nk}$,
illustrated schematically in Fig. 1, depends
on the continuum wave vector $k$
and the Landau mode index $n$. If there are no confining walls,
$W(y)=0$, one finds that
$E_n=(n+1/2)\hbar\omega_c$, $\;{\cal K}_n(k)=0$, and
$\Phi_{nk}(y)$ are the harmonic oscillator wave functions
\begin{equation}
\Phi_{nk}(y)\equiv\Phi_n(y-\ell^2k)=\frac{1}{\pi^{1/4}(2^n\ell n!)^{1/2}}
H_n\left ({y-\ell^2k\over \ell}\right )
\exp\left (-{(y-\ell^2k)^2\over 2\ell^2}\right )
\label{a5}
\end{equation}
In the presence of confining walls
the Landau-mode wave functions, $\Phi_{nk}(y)$,
are given in terms of parabolic cylinder functions. For
$\bar a\equiv a/\ell\gg n$
and $k\ell\ll \bar a$, one can use
the asymptotic expansion of parabolic cylinder
functions (Appendix A) thus obtaining
$E_n=(n+1/2+\Delta_n)\hbar\omega_c$, and
\begin{eqnarray}
{\cal K}_n(k)&=&\frac{2^n}{\sqrt{\pi} n!}
\left [(\bar a+k\ell )^{2n+1}e^{-(\bar a+k\ell )^2}+
(\bar a-k\ell )^{2n+1}e^{-(\bar a-k\ell )^2}-\Delta_n
\right ]\hbar\omega_c
\nonumber\\
&\cong &
\frac{2^{n+2}}{\sqrt{\pi}n!}{\bar a}^{2n+3}e^{-{\bar a}^2}(k\ell )^2
\hbar\omega_c,\;\;\; {\mbox {for}}\; k\ell\ll 1
\label{aa1}
\end{eqnarray}
where
$\Delta_n=2{\bar a}^{2n+1}
\exp ({-\bar a}^2)$ is the
Landau level shift due to the confining walls (this tiny shift will
be neglected in the following). One also obtains that the wave functions
$\Phi_{nk}$ are well approximated in this region
by the harmonic oscillator wave
functions: $\Phi_{nk}(y)=\Phi_n(y-\ell^2k)$.

It follows from Eqs. (\ref{a5}), (\ref{aa1}) that
the carrier velocity
$v_n=\hbar^{-1}dE_n(k)/dk\approx 0$ except for $k\sim\pm a/\ell^2$ and
$y\sim\pm a$.
Thus only the edge states provide the carriers
flow in a strong magnetic field. Classically these states correspond to
electron orbits hopping along the boundaries of a sample, Fig. 2.
Since the velocity $v_n$ is positive at the upper edge and
negative at the lower edge, the carriers can reverse
the direction of motion only by jumping to
the opposite edge\cite{but}. This process can be generated by multiple
scattering
on impurities, Fig. 2, and will be treated as follows.

Consider the general solution of Eq.\ (\ref{a1}), which can be written as
\begin{equation}
\Psi (x,y)=\phi_{nk}(x,y)+\int G(E;x,y;x',y')V(x',y')\Psi (x',y')
dx'dy'
\label{a6}
\end{equation}
where the Green's function $G$ is
\begin{equation}
G(E;x,y;x',y')=\sum_{n'}
\int\frac{e^{ik'(x-x')}\Phi_{n',k'}(y)\Phi^*_{n',k'}(y')}
{E-\left (n'+{1\over 2}\right )\hbar\omega_c-{\cal K}_{n'}(k')}dk'
\label{a7}
\end{equation}
The wave function $\phi_{nk}(x,y)$ describes the motion of carriers
inside the Landau mode $n$ and
incident from the left ($x\rightarrow -\infty$).
It follows from Eqs.\ (\ref{a6}), (\ref{a7}) that the full
solution, $\Psi (x,y)\equiv\Psi_{n,k_n} (x,y)$ at $x\rightarrow -\infty$
can be written as
\begin{equation}
\Psi_{n,k_n}(x,y)=\sum_{n'}\left [\delta_{nn'}e^{ik_nx}\Phi_{n,k_n}(y)
+(v_n/v_{n'})^{1/2}
{\mbox r}_{nn'}e^{-ik_{n'}x}\Phi_{n',-k_{n'}}(y)\right ]
\label{a8}
\end{equation}
and the same solution at $x\rightarrow +\infty$ is
\begin{equation}
\Psi_{n,k_n}(x,y)=
\sum_{n'} (v_n/v_{n'})^{1/2}
{\mbox t}_{nn'}e^{ik_{n'}x}\Phi_{n',k_{n'}}(y)
\label{a9}
\end{equation}
where $k_n$ correspond to poles in the denominator of Eq.\ (\ref{a7}):
\begin{equation}
{\cal K}_n(k_n)=E-\left (n+\frac{1}{2}\right )\hbar\omega_c
\label{a10}
\end{equation}
Here ${\mbox r}_{nn'}$ and ${\mbox t}_{nn'}$ are
the carrier reflection and transmission probability amplitudes for the
transition
from the mode $n$ into the mode $n'$. The corresponding
reflection and transmission matrices are
$R_{nn'}=|{\mbox r}_{nn'}|^2$ and
$T_{nn'}=|{\mbox t}_{nn'}|^2$. These
quantities determine the longitudinal and the Hall resistance.
In the following we consider only
the total resistance $R={\cal G}^{-1}$, where ${\cal G}$ is the
two-terminal conductance given by the
Landauer formula (in units of $e^2/h$)\cite{com1}
\begin{equation}
{\cal G}=\sum_{n,n'} T_{nn'}
\label{a11}
\end{equation}

One finds from Eq.\ (\ref{a7}) that for a given energy ($E=E_F$)
there exist two kinds of modes:
conducting modes, $n+1/2<E/\hbar\omega_c$, whose wave numbers
are real, and evanescent modes, $n+1/2>E/\hbar\omega_c$, whose
wave numbers are imaginary. The evanescent modes correspond
to virtual states and are not propagating.
Hence, $n,n' <(E/\hbar\omega_c)-1/2$
in Eq.\ (\ref{a11}). In a case of weak impurity potential $V$ the coupling
between different propagating modes is small, so that the
non-diagonal terms ($n=n'$) in the Landauer formula
can be neglected. Also
the coupling between propagating and evanescent modes is small.
As a result ${\cal G}=N$ (the total number of propagating modes
participating in the transport),
except for the Landau levels regions, $E\sim \hbar\omega_c(n+1/2)$, where
the influence of impurities is resonantly enhanced.

In the following we are going to analyze the resonant scattering near
in vicinity of Landau levels using technique of the multichannel scattering
theory.
This approach has been applied for scattering on impurities in narrow channels
with no magnetic field\cite{gl}. In the case of QHE
the multichannel treatment is more complicated, since the Landau-mode
wave functions $\Phi_{nk}(y)$ depend on the wave vector $k$.
However, the problem
can be technically simplified if we turn to the mixed,
momentum-coordinate
representation of the wave function, namely $\Psi (x,y)\rightarrow
\tilde\Psi (p,y)$, where
\begin{equation}
\tilde\Psi (p,y)=\int\Psi (x,y)e^{-ipx}dx
\label{a13}
\end{equation}
In this representation the Schroedinger equation (\ref{a1}) becomes
\begin{equation}
\left\lbrack E-\frac{1}{2m}\left (
\hbar p-{e\over c}By\right )^2
+\frac{\hbar^2}{2m}{d^2\over dy^2}-W(y)
\right\rbrack\tilde\Psi (p,y)
=\int\tilde V(p-p',y)\tilde\Psi (p',y)\frac{dp'}{2\pi}
\label{a14}
\end{equation}
with
\begin{equation}
\tilde V(p-p',y)=\int V(x,y)e^{-i(p-p')x}dx
\label{a15}
\end{equation}
and
$\phi_{nk}(x,y)\rightarrow \tilde\phi_{nk}(p,y)$ in Eq. (\ref{a2}),
which now reads
\begin{equation}
\tilde\phi_{nk}(p,y)=(2\pi )\delta (p-k_n)\tilde\Phi_n(p,y)
\label{a16}
\end{equation}
where the wave vector $k_n$ is given by Eq.\ (\ref{a10}).

In order to find the reflection and transmission amplitudes it is
useful to expand the full wave function $\tilde\Psi (p,y)$,
Eq.\ (\ref{a14}), in terms of the Landau mode wave functions $\tilde\Phi_n$
\begin{equation}
\tilde\Psi (p,y)=\sum_{n}\psi_{n}(p)\tilde\Phi_{n}(p,y)
\label{a17}
\end{equation}
Substituting Eq.\ (\ref{a17}) into Eq.\ (\ref{a14}), multiplying
it by $\Phi^*_{n'}(p,y)$ and integrating over $y$ by use of
$\langle\tilde\Phi_{np}|\tilde\Phi_{n'p}\rangle =\delta_{nn'}$,
we obtain the system
of coupled equations for the wave functions, $\psi_n$
\begin{equation}
{\cal K}_n(p)\psi_n(p)+\sum_{n'}\int V_{nn'}(p,p')
\psi_{n'}(p')\frac{dp'}{2\pi}
=(E-E_n)\psi_n(p)
\label{a18}
\end{equation}
where $E_n=(n+1/2)\hbar\omega_c$, and
\begin{equation}
V_{nn'}(p,p')=\int\tilde\Phi^*_{n'}(p',y)\tilde V(p-p',y)
\tilde\Phi_n(p,y)dy
=\sum_j\bar V_{nn'}^j(p,p')e^{-i(p-p')x_j}
\label{a19}
\end{equation}
is the impurity potential projected into the Landau-mode wave
functions space. Here
\begin{equation}
\bar V_{nn'}^j(p,p')=
\int\tilde\Phi^*_{n'}(p',y)V_j(x,y-y_j)
\tilde\Phi_n(p,y)e^{-i(p-p')x}dxdy
\label{aa19}
\end{equation}
Notice that $V_{nn'}(p,p')$ is a non-local potential,
since it depends on the momenta $p$ and $p'$ taken separately,
although the "original" impurity potential,
$V(x,y)$ (or $\tilde V(p-p',y)$, Eq. (\ref{a15}) is a local one.

Consider for the definiteness  the electron energy $E$ within
the range of $E_0\leq E\leq E_1$. Then this case only the $n$=0 Landau-mode
is the propagating mode, whereas all the other modes are evanescent.
In general, one can neglect the affect of evanescent modes on carriers
propagation
in a strong magnetic field. However, when the electron
energy $E$ approaches the nearest
evanescent Landau mode ($n=1$), its influence can be important.
We thus keep two terms, $n$=0,1 in Eq.\ (\ref{a17}), so that
the coupled equations for the first two modes are
\begin{mathletters}
\label{a20}
\begin{equation}
{\cal K}_0(p)\psi_0(p)+\int V_{00}(p,p')\psi_0(p')\frac{dp'}{2\pi}+
\int V_{01}(p,p')\psi_1(p')\frac{dp'}{2\pi}=
(E-E_0)\psi_0(p)
\label{a20a}
\end{equation}
\begin{equation}
{\cal K}_1(p)\psi_1(p)+\int V_{11}(p,p')\psi_1(p')\frac{dp'}{2\pi}+
\int V_{10}(p,p')\psi_0(p')\frac{dp'}{2\pi}=(E-E_1)\psi_1(p)
\label{a20b}
\end{equation}
\end{mathletters}
It is useful to introduce the channel-mode Green's functions
\begin{mathletters}
\label{a21}
\begin{eqnarray}
G_0(E)=(E-E_0-{\cal K}_0-V_{00})^{-1},
\label{a21a}\\
G_1(E)=(E-E_1 -{\cal K}_1-V_{11})^{-1},
\label{a21b}
\end{eqnarray}
\end{mathletters}
which describe one-dimensional motion inside the Landau modes $n=0$ and
$n=1$ respectively. Then one gets from Eqs.\ (\ref{a20b})
\begin{equation}
\psi_1=G_1V_{10}\psi_0
\label{a22}
\end{equation}
Substituting  Eq. (\ref{a22}) into Eq.\ (\ref{a20a})
one obtains the following equation for $\psi_0$
\begin{equation}
(E-E_0-{\cal K}_0-V_{00})\psi_0=V_{01}G_1(E)V_{10}\psi_0
\label{a23}
\end{equation}

Eq.\ (\ref{a23}) is the one-dimensional equation describing
a carrier propagating inside the mode $n=0$. If $V=0$ (no impurities), then
$\psi_0(p)\rightarrow 2\pi\delta (p\pm k_0)$ which corresponds to free
electron motion near the upper or the lower edge with the velocity
$v_0^{\pm}={\hbar}^{-1}\lbrack d{\cal K}_0(p)/dp\rbrack_{p=\pm k_0}$
respectively. The impurity potential $V$ generates carriers scattering. The
appropriate penetration amplitude, ${\mbox t}_{00}$,  is obtained from
$\psi_0(p)$ taken in the asymptotic regions, Eq. (\ref{a9}), which
corresponds to $p\rightarrow\pm k_0$ in the $p$-representation.
Respectively, the total conductance ${\cal G}=|t_{00}|^2$, Eq.\ (\ref{a11}).

Using the above approach we are going to investigate
resonant transitions between the edges, generated
by an interaction between carriers and
impurities. We start with a detailed analysis of the
resonant scattering on a single impurity
in the threshold regions, $E\sim E_n$.
For the definiteness we consider the electron energy within the range of
$E_0\leq E\leq E_1$, where only the Landau mode
$n=0$ is the propagating one, though the same treatment is valid in a general
case.

\section{Resonant scattering on a single impurity}

\subsection{Attractive impurity}

Consider Eq. (\ref{a23})
near the $n=1$ Landau mode threshold, $E\lesssim E_1$.
Let us assume that the operator ${\cal K}_1+V_{11}$ in the
Green's function $G_1(E)$, Eq. (\ref{a21b}),
has at least one eigenstate in the discrete spectrum, i.e.
\begin{equation}
{\cal K}_1(p)\tilde\varphi_1(p)+\int
V_{11}(p,p')\tilde\varphi_1(p')\frac{dp'}{2\pi}
=\epsilon_1\tilde\varphi_1(p)
\label{b2}
\end{equation}
where $\epsilon_1<0$. (Notice that any local one-dimensional
attractive potential generates at least one bound state.
The same is valid for the non-local potentials, $V_{nn}$).
Using the spectral representation of the Greens function
$G_1(E)$ we find for $E-E_1\rightarrow\epsilon_1$
\begin{equation}
G_1(E)=\frac{|\tilde\varphi_1\rangle\langle\tilde\varphi_1 |}
{E-E_1-\epsilon_1}
\label{b3}
\end{equation}
Then Eq.\ (\ref{a23}) is reduced to
\begin{equation}
(E-E_0-{\cal K}_0-V_{00})\psi_0=\hat V\psi_0
\label{b4}
\end{equation}
where the potential
$\hat V$ is the energy dependent and has the separable form
\begin{equation}
\hat V(E,p,p')=
\frac{\langle p|V_{01}|\tilde\varphi_1\rangle\langle\tilde
\varphi_1 |V_{10}|p'\rangle}
{E-E_1-\epsilon_1 }
\label{b5}
\end{equation}

Let us consider weak impurity potential,
$|V_{00}|\ll E_1-E_0=\hbar\omega_c$. In this case one can neglect $V_{00}$ in
Eqs. (\ref{a21a}),(\ref{b4}), which become
\begin{equation}
(E-E_0-{\cal K}_0)\psi_0=\hat V\psi_0
\label{bb2}
\end{equation}
\begin{equation}
G_0(E;p,p')=\frac{2\pi\delta (p-p')}{E-E_0 -{\cal K}_0(p)}
\label{b8}
\end{equation}
However, the impurity potential in the r.h.s. of Eq. (\ref{bb2})
cannot be neglected. It follows from Eq. (\ref{b5}) that
$\hat V\rightarrow\infty$ when $E-E_1\rightarrow\epsilon_1$ for no matter
how weak the impurity is.  Therefore
one cannot use any finite order Born approximation for
Eq. (\ref{bb2}), but rather the whole Born series
\begin{equation}
|\psi_0\rangle =|k_0\rangle +G_0\hat V|k_0\rangle +G_0\hat VG_0\hat V
|k_0\rangle +\cdots =(1+G_0t)|k_0\rangle ,
\label{b6}
\end{equation}
where $\langle p|k_0\rangle =2\pi\delta (p-k_0)$, and
$t$ is the scattering operator, which satisfies the
Lippmann-Schwinger equation
\begin{equation}
t=\hat V +\hat VG_0t
\label{b7}
\end{equation}
The wave vector $k_0$ is defined from the equation
\begin{equation}
{\cal K}_0(k_0)=E-E_0
\label{bb1}
\end{equation}
Since $E-E_0\approx\hbar\omega_c$, one finds that
$k_0\ell^2\sim a\gg \ell$ (see Fig. 1).

Substituting $\psi_0(p)\equiv\langle p|\psi_0\rangle$, given by
Eq. (\ref{b6}) into Eq. (\ref{a17}) for $n=0$ and
using Eq. (\ref{b8}) we get
\begin{equation}
\tilde\Psi_{0,k_0}(p,y)=\left [ 2\pi\delta (p-k_0)+\frac{\langle p|t|k_0
\rangle}{E-E_0 -{\cal K}_0(p)}\right ]\tilde \Phi_{0,k_0}(p,y)
\label{b9}
\end{equation}
The inverse Fourier transform
$\tilde\Psi_{0,k_0}(p,y)\rightarrow \Psi_{0,k_0}(x,y)$ gives rise for
the total wave function $\Psi_{0,k_0}(x,y)$ at $x\rightarrow +\infty$
\begin{equation}
\Psi_{0,k_0}(x,y)=e^{ik_0x}\Phi_{0,k_0}(y)+\frac{1}{i\hbar v_0}
\langle k_0|t|k_0\rangle e^{ik_0x}\Phi_{0,k_0}(y)
\label{b10}
\end{equation}
where $v_0={\hbar}^{-1}\lbrack d{\cal K}_0(p)/dp\rbrack_{p=k_0}$
is the carriers velocity inside the $n=0$ Landau mode. As a result, the
corresponding transmission coefficient is
\begin{equation}
T_{00}=\left |1+\frac{1}{i\hbar v_0}\langle k_0|t|k_0\rangle\right |^2
\label{b11}
\end{equation}
where the scattering operator
$t$ is given by the Lippmann-Schwinger equation
(\ref{b7}). Since the potential
$\hat V$ in Eq. (\ref{b5}) is a separable
one, the Lippmann-Schwinger equation can be easily solved by taking
$\langle p|t|p'\rangle\equiv t(E,p,p')$ in a form
\begin{equation}
t(E,p,p')=\frac{\langle p|V_{01}
|\tilde\varphi_1\rangle\langle\tilde\varphi_1 |V_{10}|p'\rangle}{D(E)}
\label{b12}
\end{equation}
Substituting Eq. (\ref{b12}) into Eq. (\ref{b7}) and using
Eq. (\ref{b5}), we find that
\begin{equation}
D(E)=E-E_1-\epsilon_1-\langle \tilde\varphi_1|V_{10}G_0V_{01}|\tilde\varphi_1
\rangle
\label{b13}
\end{equation}
Using Eq. (\ref{b8}) for $G_0$ one obtains
\begin{equation}
{\mbox {Im}}\ \langle \tilde\varphi_1|V_{10}G_0V_{01}|\tilde\varphi_1\rangle
=-(\tilde\Gamma_1^++\tilde\Gamma_1^-)
\label{b14}
\end{equation}
where
\begin{equation}
\tilde\Gamma_1^{\pm}=\frac{1}{2\hbar v_0}\left |
\langle\pm  k_0|V_{01}|\tilde\varphi_1\rangle\right |^2
\label{b15}
\end{equation}
are the partial widths which acquires the bound state in the evanescent
mode $n$=1 due to decay into the upper or into the lower edge states of
the propagating mode $n$=0.

Substituting Eq. (\ref{b12}) into Eq. (\ref{b11}) and using
Eqs. (\ref{b13}) -- (\ref{b15}) we find
\begin{equation}
T_{00}=\left |1-\frac{2i\tilde\Gamma_1^+}{E-E_1
-\epsilon_1-\tilde\Delta_1 +
i(\tilde\Gamma_1^++\tilde\Gamma_1^-)}\right |^2
\label{b16}
\end{equation}
where $\tilde\Delta_1 ={\mbox {Re}}\ \langle
\tilde\varphi_1|V_{10}G_0V_{01}|\tilde\varphi_1\rangle$
is the shift of the bound state energy due to decay to the edge states
(usually $\tilde\Delta_1\ll\epsilon_1$).
Finally we obtain for the two-terminal conductance
\begin{equation}
{\cal G}(E)=1- \frac{4\tilde\Gamma_1^+\tilde\Gamma_1^-}{
(E-\tilde E_1)^2+(\tilde\Gamma_1^++\tilde\Gamma_1^-)^2}
\equiv 1-{\cal R}(E)
\label{b17}
\end{equation}
where $\tilde E_1=E_1+\epsilon_1+\tilde\Delta_1$, and
\begin{equation}
{\cal R}(E)=|t(E,k_0,-k_0)|^2/\hbar^2v_0^2
\label{ab17}
\end{equation}
is the reflection coefficient. The latter is the probability
of the inter-edge transition across a sample.

In general, when the electron energy approaches the threshold of the $N$'
Landau mode, the same approach results in
\begin{equation}
{\cal G}=N- \frac{4\tilde\Gamma_N^+\tilde\Gamma_N^-}{
(E-\tilde E_N )^2+(\tilde\Gamma_N^++\tilde\Gamma_N^-)^2}
\label{bb17}
\end{equation}
where the expressions for $\tilde\Gamma_N^{\pm}$
and $\tilde E_N$ are completely analogous to those for $N=1$.
Similar result for the total conductance ${\cal G}$
in a case of short range impurity  has been obtained
in Ref.\cite{suk} by using a different technique.

We thus demonstrated that any weak attractive impurity generates the
resonance dip in the total conductance (the peak in the
longitudinal resistance).
This dip (peak) appears as a result of the
backward scattering due to inter-edge transition via
the resonance state in the nearest evanescent mode.
This inter-mode scattering mechanism
is typical for an attractive impurity which
produces bound states below thresholds of the Landau modes.

\subsection{Attractive short-range impurity}

Now we apply the results obtained above for the evaluation  of
the resonance energy and the partial widths in a case of an
attractive short range impurity potential, i.e. the interaction range
is much smaller than the magnetic length $\ell$. In this case one
can write for $V_{nn'}$, Eq. (\ref{a19}),
\begin{eqnarray}
V_{nn'}(p,p') &=& \int\tilde\Phi^*_{n'}(p',y)V(x-x_i,y-y_i)
\tilde\Phi_n(p,y)e^{-i(p-p')x}dxdy
\nonumber\\
&\simeq& \lambda\Phi_n (y_i-\ell^2p)\Phi_{n'} (y_i-\ell^2p')
\label{c1}
\end{eqnarray}
where  $\lambda =\int V(x,y)dxdy<0$ and
$\Phi_n$ are the Harmonic oscillator wave functions, Eq. (\ref{a5}).
The impurity coordinate $x_i$ is chosen zero.
Here we assumed that that
the impurity is not in the edge regions, so
that the Landau-mode wave functions can be well approximated by the harmonic
oscillator functions, $\tilde\Phi_n(p,y)=
\Phi_n (y_i-\ell^2p)$.

Since the potential $V_{11}$, Eq. (\ref{c1}) has a separable form,
the Schroedinger equation (\ref{b2}) for the bound state can be easily solved.
One obtains
\begin{equation}
\tilde\varphi_1(p)={\cal N}
\frac{\Phi_1(y_i-\ell^2p)}{\epsilon_1 -{\cal K}_1(p)}
\label{c2}
\end{equation}
where ${\cal N}$ is the normalization:
\begin{equation}
{\cal N}^{-2}=\int
\frac{\Phi_1^2(y_i-\ell^2p)}{[\epsilon_1 -{\cal K}_1(p)]^2}
\frac{dp}{2\pi}
\label{c3}
\end{equation}
and the bound state energy $\epsilon_1$ is defined by the equation
\begin{equation}
1-\lambda\int\frac{\Phi_1^2(y_i-\ell^2p)}{\epsilon_1 -{\cal K}_1(p)}
\frac{dp}{2\pi}=0
\label{c4}
\end{equation}
Since ${\cal K}_1(p)\geq 0$, it is clear that for
the attractive interaction
($\lambda <0$), and no matter how small the $\lambda$ is, there always
exists $\epsilon_1<0$, which satisfies Eq. (\ref{c4}).
One can easily evaluate $\epsilon_1$ and ${\cal N}$
by taken into account that
${\cal K}_1\cong 0$ in the region of $p$ which makes the dominant
contribution in the integrals (\ref{c3}), (\ref{c4}). Then
\begin{equation}
\epsilon_1=\frac{\lambda}{2\pi\ell^2},\;\;\;\;\;\;\;
{\cal N}=(\lambda\epsilon_1)^{1/2}
\label{cc4}
\end{equation}

The bound state $\tilde\varphi_1$ which appears in the evanescent mode ($n=1$),
generates the resonance in the propagating mode ($n$=0)
through the mode-mixing  potential $V_{01}$, Eq. (\ref{c1})
\begin{equation}
V_{01}(p,p')=\lambda\Phi_0 (y_i-\ell^2p)\Phi_1 (y_i-\ell^2p')
\label{c5}
\end{equation}
Substituting Eq. (\ref{c5}) into Eq. (\ref{b15}) and using
Eqs. (\ref{a5}) and (\ref{cc4})
we find for the partial widths
\begin{equation}
\tilde\Gamma_1^{\pm}
=\frac{\epsilon_1^2\sqrt{\pi}\ell}{\hbar\tilde v_0}\exp
\lbrack -(y_i\mp \ell^2\tilde k_0)^2/\ell^2\rbrack
\label{c6}
\end{equation}
where $\tilde k_0$ is the electron wave vector
at the resonance energy, which is defined from the equation
${\cal K}_0(\tilde k_0)=E_1-E_0-\epsilon_1$, and the velocity
$\tilde v_0=\hbar^{-1}\lbrack d{\cal K}_0(p)/dp\rbrack_{p=\tilde k_0}$.
One can estimate from Fig. 1b that
$\tilde k_0\ell^2\sim a$ and $\tilde v_0\sim \ell\omega_c$.

It follows from Eq. (\ref{c6}) that $\tilde\Gamma_1^+/\tilde\Gamma_1^-
\sim\exp (4y_ia/\ell^2)$. Therefore for a wide sample ($a/\ell\gg 1$)
the resonance partial widths
$\tilde\Gamma_1^+$ and $\tilde\Gamma_1^-$ are largely different
unless $4y_ia\sim\ell^2$.
As a result the depth of the resonant dip
in the total conductance, Eq. (\ref{b17}), is exponentially small
\begin{equation}
(\Delta {\cal G})_{max}={\cal R}(\tilde E_1)=
\frac{4\tilde\Gamma_1^+/\tilde\Gamma_1^-}
{(\tilde\Gamma_1^+/\tilde\Gamma_1^-+1)^2}\sim 4e^{-4\tilde k_0y_i}
\label{bbb17}
\end{equation}
and therefore the resonance is strongly suppressed,
unless the impurity is not in the middle of a sample\cite{suk}. Then
the resonance peak appears in the longitudinal resistance, although its width
would be exponentially small.

\subsection{Short range repulsive impurity}

Using classical mechanics arguments one finds that a
repulsive impurity potential
in the presence of strong magnetic field
can trap electrons moving in a confining conducting channel.
However, in quantum mechanical sense, such a state is not
a bound one. It is rather a quasi-bound (resonance) state because
the energy spectrum of the corresponding Hamiltonian, Eq. (\ref{a1}),
is continuous and there are no normalized eigenstates exist in
continuum spectrum.  Similar to the previous case such a quasi-bound state
can generate resonant transitions of carriers between the
edges. The difference is that the resonances from repulsive impurities appear
above thresholds of the propagating Landau modes,
whereas the resonances generated by attractive impurities are below
thresholds. As a result repulsive impurities can produce direct
inter-edge transitions inside the propagating modes
(the inner-mode transitions), in a contrast with repulsive impurities, which
generate inter-edge transitions via bound states in the
evanescent modes, (the inter-mode transitions).

Let us investigate these resonances using the same couple-channel technique.
Consider the electron energy $E$ in vicinity of
the threshold of the $n=0$ Landau mode, $E\sim E_0$. If the impurity is weak,
$V_{01}G_1(E)V_{01}\sim |V_{00}|^2/(E_1-E_0)\ll V_{00}$, the
right-hand side term in Eq. (\ref{a23}) can be neglected. Then this equation
can be rewritten as
\begin{equation}
(E-E_0-{\cal K}_0 -V_{00})\psi_0=0
\label{b19}
\end{equation}
Notice that $V_{00}$ is a {\em non-local} one-dimensional repulsive potential.
Therefore it can generate resonances, though a {\em local}
one-dimensional {\em repulsive} potential cannot.

We start with a short range repulsive impurity. Then
the potential $V_{00}$, Eq. (\ref{c1}), has the same separable form
as in the case of an attractive short range impurity:
\begin{equation}
V_{00}(p,p')=\lambda g_0(p)g_0(p')
\label{b18}
\end{equation}
where  $\lambda =\int V(x,y)dxdy>0$ and
$g_0(p)=\Phi_0 (y_i-\ell^2p)$.
The corresponding transmission coefficient is given by Eq. (\ref{b11}),
with the scattering operator $t$
obtained from the Lippmann-Schwinger equation
\begin{equation}
t=V_{00}(1+G_0t)
\label{bb19}
\end{equation}
and the reflection coefficient, ${\cal R}(E)$, i.e.
the probability of the inter-edge transition, is given by Eq. (\ref{ab17}).
Taking $t(E,p,p')$ in a form
\begin{equation}
t(E,p,p')=\lambda g_0(p)g_0(p')/D(E)
\label{b20}
\end{equation}
we get
\begin{equation}
D(E)=
1-\int\frac{\lambda g_0^2(p'')}{E-E_0-{\cal K}_0(p'')+i\delta}
\frac{dp''}{2\pi}
\label{b21}
\end{equation}
Since $\lambda$ is positive, one easily finds that
Re$[D(E)]$ vanishes for $E=E_0+\epsilon_0$, where
$\epsilon_0>0$ is defined from the equation
\begin{equation}
1-{\cal P}\int\frac{\lambda g_0^2(p'')}{\epsilon_0 -{\cal K}_0(p'')}
\frac{dp''}{2\pi}=0
\label{b22}
\end{equation}
Here ${\cal P}$ denotes the principal value of the integral.
Hence, one gets for $E-E_0\sim\epsilon_0$,
\begin{equation}
D(E)={\cal A}\left [E-E_0-\epsilon_0 +i(\Gamma_0^++
\Gamma_0^-)\right ]
\label{bbb2}
\end{equation}
where ${\cal A}=d[{\mbox {Re}}D(E)]/dE|_{E=E_0+\epsilon_0}$ and
$\Gamma_0^{\pm}$  are obtained from the singular part of Eq.(\ref{b21}).
As in the previous case of attractive impurity
one can evaluate the resonance energy $\epsilon_0$ and ${\cal A}$
by taken ${\cal K}_0\sim 0$ in Eq. (\ref{b22}), thus obtaining
\begin{equation}
\epsilon_0=\lambda\int g_0^2(p'')\frac{dp''}{2\pi}=\frac{\lambda}
{2\pi\ell^2}
\label{bb22}
\end{equation}
and ${\cal A}=\epsilon_0^{-1}$, where the partial widths are given by
\begin{equation}
\Gamma_0^{\pm}=\frac{\lambda}{2\bar v_0}{\cal A}^{-1}
g_0^2(\pm \bar k_0)=
\frac{\epsilon_0^2\sqrt{\pi}\ell}{\hbar\bar v_0}
\exp\lbrack -(y_i\mp \ell^2\bar k_0)^2/\ell^2\rbrack
\label{b23}
\end{equation}
Here $\bar k_0$ is determined from the equation
${\cal K}_0(\bar k_0)=\epsilon_0$, and
$\bar v_0={\hbar}^{-1}\lbrack d{\cal K}_0(p)/dp\rbrack_{p=\bar k_0}$
is the carrier velocity.

Substituting Eq. (\ref{bbb2}) into Eq. (\ref{b20}) we find that
the scattering amplitude $t(E,p,p')$
has the Breit-Wigner form, which means the existence of the
resonance above the Hall plateau at the energy
$\bar E_0=E_0+\epsilon_0$ and of the width
$\Gamma_0^++\Gamma_0^-$. Using Eq. (\ref{b11})
we finally obtain for the two-terminal conductance ${\cal G}$ in vicinity of
the resonance: ${\cal G}=1-{\cal R}$, where
\begin{equation}
{\cal R}(E)=\frac{4\Gamma_0^+\Gamma_0^-}{
(E-\bar E_0 )^2+(\Gamma_0^++\Gamma_0^-)^2}
\label{b25}
\end{equation}

We thus found that both repulsive and attractive
impurities produce the resonance dip in
the total conductance (and therefore the peak in the
longitudinal resistance). It both cases the resonance energy and the width
are given by the same expressions,
Eqs. (\ref{b23}),(\ref{bb22}), and Eqs. (\ref{cc4}),
(\ref{c6}), respectively, though the resonances
appear near different Landau
modes. Notice that the carrier velocity  $\bar v_0$
in Eq. (\ref{b23}) is smaller than the corresponding velocity
$\tilde v_0$ entering Eq. (\ref{c6}) (see Fig. 1a). As a result
the resonance generated by
a single repulsive impurity is more broadened than that from a single
attractive impurity. However,
the mostly pronounced distinction between these two
cases takes place for extremely weak impurities,
$\lambda\rightarrow 0$. Then
the values of the corresponding wave vectors, $\tilde k_0$ and $\bar k_0$,
are very different. The wave vector
$\tilde k_0$ for the attractive impurity is always large,
$\tilde k_0\ell\sim a/\ell\gg 1$, even when
$\lambda\rightarrow 0$ (see  Fig. 1a). In a contrast, the wave vector
$\bar k_0\rightarrow 0$ for $\lambda\rightarrow 0$. Indeed, one obtains from
Eq. (\ref{aa1}) for $\epsilon_0/\hbar\omega_c\sim \exp (-\bar a^2)$
\begin{equation}
\bar k_0\ell =\left (\frac{\sqrt{\pi}}{4\bar a^3}e^{\bar a^2}
\frac{\epsilon_0}{\hbar\omega_c}\right )^{1/2}=\left (\frac{\lambda\ell}
{8a^3\sqrt{\pi}\hbar\omega_c}\right )^{1/2}e^{a^2/2\ell^2}
\label{bb25}
\end{equation}
Correspondingly, the partial widths, Eq. (\ref{b23}), are
\begin{equation}
\Gamma_0^{\pm}=\frac{\epsilon_0^{3/2}}{(\hbar\omega_c)^{1/2}}
\frac{\pi^{3/4}}{4\bar a^{3/2}}e^{\bar a^2/2}
\exp\left [-\left (\frac{y_i}{\ell}\mp \bar k_0\ell \right)^2\right]
\label{d9}
\end{equation}
Therefore, $\Gamma_0^+\sim \Gamma_0^-$, irrespectively on the position
of the impurity $y_i$, since $\bar k_0\ell\ll 1$. As a result
the resonance
dip of total conductance, Eq. (\ref{b25}),
may survive even in a wide sample.
Such a different behavior of partial widths for
attractive and repulsive impurities is easy to interpret by notice that the
very weak attractive impurity generates the
resonance just above the Hall plateau, where the
corresponding wave vector is small.
In this case the effective electron spreading is large and therefore
the resonant scattering would be insensitive to the impurity
position.

\subsection{General case}
If the repulsive impurity potential, $V(x,y)$,
is not a short range one, then the corresponding non-local
potential $V_{00}$ in Eq. (\ref{b19}) cannot be written in the separable form
of Eq. (\ref{b18}). In this case our previous treatment cannot be applied.
However, one can benefit by the
effective kinetic term ${\cal K}_0(p)$ is almost zero except for
$p\sim \pm a/\ell^2$. That allows to consider the kinetic term
as a perturbation. Let us put ${\cal K}_0=0$ in Eq. (\ref{b19})
(the infinite mass approximation). In this case
the resonance state $\psi_0(p)$,
generated by Eq. (\ref{b19}) appears as a stable
state described by the normalized wave function,
$\psi_0(p)\rightarrow \varphi_0(p)$, and
Eq. (\ref{b19}) can be rewritten as
\begin{equation}
\int V_{00}(p,p')\varphi_0 (p')\frac{dp'}{2\pi}=\epsilon_0\varphi_0(p)
\label{ab19}
\end{equation}
Here $\epsilon_0=E-E_0$ is the resonance energy.

When the neglected kinetic term ${\cal K}_0$ is "turned on",
the bound state $\varphi_0$ gets the width, because
${\cal K}_0$ provides the coupling with continuum of the edge
states. In order to calculate the partial widths we apply the
approach developed in\cite{gl,gur} for treatment of the
resonance states. We found in\cite{gl} that
the corresponding partial widths are obtained in terms of the matrix elements
of the impurity potential between the bound state and
the free (continuum) states
\begin{equation}
\Gamma_0^{\pm}=\frac{1}{2\hbar\bar v_0}|<\varphi_0|V_{00}|\mp\bar k_0>|^2=
\frac{\epsilon_0^2}{2\hbar\bar v_0}|<\varphi_0|\mp\bar k_0>|^2
\label{bb20}
\end{equation}
where Eq. (\ref{ab19}) has been used in order to eliminate the impurity
potential
$V_{00}$. Correspondingly, the scattering amplitude $t$, Eq. (\ref{bb19}),
in the vicinity of the resonance is given by the Breit-Wigner formula
\begin{equation}
t(E,p,p')=\frac{\epsilon_0^2<p|\varphi_0><\varphi_0|p'>}{E-E_0-\epsilon_0+
i(\Gamma_0^++\Gamma_0^-)}
\label{bb10}
\end{equation}
Finally, the probability of the inter-edge transition (the reflection
coefficient)
is determined by Eq. (\ref{ab17}), namely ${\cal R}=|t(E,\bar k_0,-\bar
k_0)|^2/
\hbar^2\bar v_0^2$.

As an example, let us apply Eqs. (\ref{ab19}), (\ref{bb20}) for
a short range repulsive potential. Substituting Eq. (\ref{b18}) into
Eq. (\ref{ab19}) one finds that
the solution is
\begin{equation}
\varphi_0(p) =C\,g_0(p)
\label{bb23}
\end{equation}
where $C=(2\pi\ell )^{1/2}$
is the normalization coefficient, and the resonance energy
$\epsilon_0$ is given by Eq. (\ref{bb22}). Also, substituting
Eq. (\ref{bb23}) into
Eq. (\ref{bb20}) we reproduce Eq. (\ref{b23}) for the partial widths.
Notice that Eq. (\ref{ab19}) has no solutions if
$V_{00}$ is a local potential: $V_{00}\equiv V_{00}(p-p')$.

\subsection{Repulsive impurity of Gaussian-type}

Let us consider an another example of the carrier-impurity interaction,
namely the Gaussian potential
\begin{equation}
V(x,y)=\frac{\lambda}{\pi r_0^2}\exp [-
({\mbox{\boldmath $r$}}-{\mbox{\boldmath $r$}_i})^2/r_0^2]
\label{e1}
\end{equation}
where
${\mbox{\boldmath $r$}}=(x,y)$, and ${\mbox{\boldmath $r$}}_i=(x_i,y_i)$.
The interaction range $r_0$ is taken as a free parameter. Hence, $V(x,y)$
can describe either short or long-range impurity.
We restrict ourself with a repulsive interaction ($\lambda >0$), although our
treatment can be easily performed for an attractive interaction as well.

Using Eqs. (\ref{a5}), (\ref{a19}) and taking $x_i=0$
we get for the potential $V_{00}$ the following expression:
\begin{equation}
V_{00}(p,p')=\frac{\lambda}{\sqrt{\pi} R_0}
\exp \left [ -\frac{\ell^2}{R_0^2}\left (\frac{y_i}{\ell}-\ell
\frac{p+p'}{2}\right )^2-\frac{R_0^2}{4}(p-p')^2\right ]
\label{e2}
\end{equation}
where $R_0=(r_0^2+\ell^2)^{1/2}$.
Notice that for $r_0\ll\ell$ (the short range interaction)
\begin{equation}
V_{00}(p,p')\rightarrow\frac{\lambda}{\sqrt{\pi}\ell}
e^{-(y_i-\ell^2p)^2/2\ell^2}e^{-(y_i-\ell^2p')^2/2\ell^2}
\label{e3}
\end{equation}
i.e. the potential obtains the separable form of Eq.(\ref{b18}).
However, in a case of long range interaction, $r_0\gg\ell$, the potential turns
to a local one:
\begin{equation}
V_{00}(p,p')\rightarrow\frac{\lambda}{\sqrt{\pi}r_0}
e^{-[4y_i^2+r_0^4(p-p')^2]/4r_0^2}
\label{e4}
\end{equation}

Substituting Eq. (\ref{e2}) into
Eq. (\ref{ab19}) and using the variables
$q=y_i-\ell^2 p$ and $q'=y_i-\ell^2 p'$ instead of $p$ and $p'$ we get
\begin{equation}
\int\tilde V_{00}(q,q')\varphi_0(q')\frac{dq'}{2\pi\ell^2}=
\epsilon_0\varphi_0(q)
\label{e5}
\end{equation}
where
\begin{equation}
\tilde V_{00}(q,q')=\frac{\lambda}{\sqrt{\pi}R_0}
e^{-\alpha (q^2+{q'}^2)+2\beta qq'}
\label{e6}
\end{equation}
Here the parameters $\alpha$ and $\beta$ are
\begin{equation}
\alpha =\frac{R_0^4+\ell^4}{4 R_0^2\ell^4},~~~~~~\beta=
\frac{R_0^4-\ell_0^4}{4 R_0^2\ell^4}
\label{e7}
\end{equation}
One finds that the wave function $\varphi_0$ taken in a
Gaussian form
\begin{equation}
\varphi_0(q)=2^{3/4}(\pi\eta )^{1/4}\ell\exp (-\eta q^2),
\label{ee7}
\end{equation}
satisfies Eq. (\ref{e5}) for
$\eta =1/2\ell^2$, where the resonance energy is
\begin{equation}
\epsilon_0=\frac{\lambda}{\pi (R_0^2+\ell^2)}
\label{e8}
\end{equation}
Using Eqs. (\ref{bb20}) and (\ref{ee7}) we obtain for the partial widths
the following result
\begin{equation}
\Gamma_0^{\pm}=\frac{\epsilon_0^2\sqrt{\pi}\ell}{\hbar\bar v_0}
\exp\lbrack -(y_i\mp \ell^2\bar k_0)^2/\ell^2\rbrack
\label{e9}
\end{equation}
which coincides with Eq. (\ref{b23}) for a short range impurity.

Comparing $\varphi_0(q)$ given by  Eq. (\ref{ee7}) with the wave function for a
short range impurity, Eq. (\ref{bb23}), we find that
the both wave functions coincide. It is
rather remarkable that the range of the Gaussian potential does
not enter into $\varphi_0$, so that
the wave function spreading is the same as for a
short range impurity. The interaction range $r_0$ appears only in the resonance
energy, Eq. (\ref{e8}). One finds that in the
limit of $r_0\rightarrow\infty$ the energy  $\epsilon_0 \rightarrow 0$, and
the resonance disappears. This result is rather expectable
since the potential $V_{00}$ in this limit turns to be a
local repulsive potential, Eq. (\ref{e4}), which does not produces
any quasi-stationary states.

\section{Resonant scattering on two impurities}

It was found in the previous section that the
inter-edge resonant transport via single impurity
drops down very rapidly with a sample width.
Let us consider an another mechanism of inter-edge transport, when
a carrier reaches the opposite edge through
subsequent resonant hopping via many impurities.
We start with an example of a such resonant
hopping via two identical impurities.
In this case the potential $V(x,y)$ in Eq. (\ref{a1}) is
\begin{equation}
V(x,y)=V(x-x_1,y-y_1)+V(x-x_2,y-y_2)
\label{f1}
\end{equation}
For the definiteness we discuss two
repulsive impurities, although the analysis is valid for attractive
impurities as well.

Consider  Eq. (\ref{b19}) near the $n$=0 Landau mode threshold.
Using Eq. (\ref{a19}) one can rewrite the potential $V_{00}$ as
\begin{equation}
V_{00}(p,p')=\bar V_{00}^{(1)}(p,p')e^{-i(p-p')x_1}+
\bar V_{00}^{(2)}(p,p')e^{-i(p-p')x_2}
\label{f2}
\end{equation}
It follows from our previous analysis that each of impurity potentials in
Eq. (\ref{f2}) generates the resonance above the Hall plateau at the same
energy, $\bar E_0=E_0+\epsilon_0$, where $\epsilon_0=\lambda /2\pi\ell^2$,
Eq. (\ref{bb22}). Neglecting the kinetic term  ${\cal K}_0$ in
Eq. (\ref{b19}), one finds that each of the resonances becomes
stable state, described by the normalized
wave functions
\begin{equation}
\varphi_j(p)=\bar\varphi_j(p)\exp (-ipx_j)
\label{f3}
\end{equation}
where $j=1,2$ and $\bar\varphi_j(p)$ is obtained from Eq. (\ref{ab19}) for
$V_{00}(p,p')=\bar V_{00}^j(p,p')$.
Notice that the phase factor in Eq. (\ref{f3}) would
play now an important role, although it was insignificant in the single
scattering
process. Similar to the previous case, the states
$\varphi_{j}(p)$ obtain the widths $\Gamma_{j}=
\Gamma_{j}^++\Gamma_{j}^-$,
when the neglected kinetic term ${\cal K}_0$ is "switched on".
Here $\Gamma_j^{\pm}$ denote the partial widths due to decay into the upper
and the lower edge state respectively, Eq. (\ref{b23}).

One can easily realize that the resonance transport of carriers via
two impurities is completely analogous to the
resonant tunneling through
one-dimensional double-well potential system with aligned levels.
This process has been studied in details by using different techniques, see
for instance\cite{lm,sok,gur1} and references therein.
Here we adopt the time-dependent approach of
Ref.\cite{gur1}, since it is easily extended for a case
of many impurities. According to this approach the amplitude of
resonant scattering on two impurities can be obtained
directly from
the Schroedinger equation $i\hbar \dot\psi (t)={\cal H}\psi (t)$
for two states wave function $\psi (t)=(b_1(t),b_2(t))$, with the
initial condition $\psi (0)=(1,0)$, where
$b_{1,2}(t)$ are the probability amplitudes
to find the carrier in the states $\varphi_{1,2}$ respectively, and
${\cal H}$ is the effective Hamiltonian
\begin{equation}
{\cal H}=\left (\matrix{\bar E_0-i\Gamma_1&\Omega_{21}\cr
\Omega_{12}&\bar E_0-i\Gamma_2\cr}\right )
\label{f4}
\end{equation}
The off-diagonal matrix element  $\Omega_{12}$
is the hopping transition amplitude between
the states, $\varphi_{1,2}$. The diagonal energy shifts,
$\Delta_{1}=\Delta_{2}\ll |\Omega_{12}|$ (see Appendix B) were included in the
energy $\bar E_0$. The amplitude $b_2$ determines the
scattering amplitude Eq. (\ref{bb19})
in a vicinity of the resonance, $E\sim \bar E_0=E_0+\epsilon_0$ by means of
\begin{equation}
\hbar \ t(E,p,p')=<p|V_{00}^{(1)}|\varphi_1>\tilde b_2(E)
<\varphi_2|V_{00}^{(2)}|p'>=
\epsilon_0^2<p|\varphi_1>\tilde b_2(E)<\varphi_2|p'>
\label{f5}
\end{equation}
(c.f. with Eq. (\ref{bb10})), where $\tilde b_j$ is the Laplace transform
of $b_j$:
\begin{equation}
\tilde b_j(E)=\int_0^{\infty}\exp (iEt/\hbar)b_j(t)dt
\label{f6}
\end{equation}

Solving these equations we obtain for the probability of the inter-edge
transition,
${\cal R}(E)$, Eq. (\ref{ab17}),
the following result
\begin{equation}
{\cal R}(E)=\frac{4\Gamma_1^+|\Omega_{12}|^2\Gamma_2^-}{|\det [E-{\cal H}]|^2}
\label{f7}
\end{equation}
where
\begin{equation}
\det [E-{\cal H}]=(E-\bar E_0+i\Gamma_1)(E-\bar E_0+i\Gamma_2)
-|\Omega_{12}|^2
\label{f8}
\end{equation}
Thus, ${\cal R}(E)$ coincides with
the probability of the resonance passage through a double-well potential
for aligned levels\cite{lm,sok}.
One finds from Eqs. (\ref{f7}),( \ref{f8}) that ${\cal R}(E)$
has two peaks near the Landau levels at the energies
\begin{equation}
E_{\pm}=\bar E_0\pm \sqrt{|\Omega_{12}|^2-\frac{(\Gamma_1-\Gamma_2)^2}{4}}
\label{ff7}
\end{equation}
which reflects the splitting of the resonance due to off-diagonal transitions
between the two impurities. The reflection coefficient ${\cal R}(E)$
reaches the maxima at $E=E_{\pm}$
\begin{equation}
{\cal R}_{max}=\frac{4\Gamma_1^+\Gamma_2^-}{(\Gamma_1^++\Gamma_2^-)^2}
\frac{\displaystyle|\Omega_{12}|^2}{\displaystyle
{|\Omega_{12}|^2+\frac{(\Gamma_1^++\Gamma_2^-)^2}{16}}}
\label{ff8}
\end{equation}
It is quite clear from Eq. (\ref{ff8}) that the inter-edge transition via two
impurities is more likely than the direct transition via
one impurity. For instance, it is not required for the impurities
to be in the middle of a sample in order to generate non-vanishing
resonance peak in ${\cal R}(E)$.

Using Eq. (\ref{bp11}) from Appendix B for ${\cal K}_0=0$, and
Eqs. (\ref{bb23}) for the wave functions $\varphi_{1,2}$,
in a case of short range (or Gaussian) impurity potentials,
we obtain for the hopping matrix element
\begin{equation}
\Omega_{ij} =\epsilon_0\int\varphi_i(p)\varphi_j^*(p)
\frac{dp}{2\pi}=\epsilon_0
e^{-({\mbox{\boldmath $r$}_i}-{\mbox{\boldmath $r$}_j})^2/4\ell^2}
e^{-i(x_i-x_j)(y_i+y_j)/2\ell^2}
\label{f9}
\end{equation}
where for the diagonal energy shifts $\Delta_1=\Delta_2$
one gets (for a short range potential)
\begin{equation}
\Delta_i =\int\varphi_i(p)V_{00}^j(p,p')\varphi_i^*(p')
\frac{dpdp'}{4\pi^2}=\epsilon_0
e^{-({\mbox{\boldmath $r$}_i}-{\mbox{\boldmath $r$}_j})^2/2\ell^2}
\label{ff9}
\end{equation}
As expected, $\Delta_i \ll |\Omega_{ij}|$.

It is important to point out that the non-diagonal hopping elements are
complex,
$\Omega_{ij}\equiv |\Omega_{ij}|\exp(-i\alpha_{ij})$,
with the phase
\begin{equation}
\alpha_{ij}=\pi B(x_i-x_j)(y_i+y_j)/\phi_0
\label{f10}
\end{equation}
where $\phi_0=hc/e$ is the flux quantum.

\section{Resonant scattering on $n$ impurities}

\subsection{Interference effects}

Our results for resonant scattering on two impurities can be
extended for any number of impurities inside a sample.
Let us consider the transport of a carrier from the
upper to the lower edge as a sequence of resonant transitions
via $n$ impurities located at the points ${\mbox{\boldmath $r$}_1},\ldots,
{\mbox{\boldmath $r$}_n}$ inside a sample.
We enumerate impurities in such a way that the first one is closest to the
upper edge, and the last one is closest to the lower edge.
As in the previous case we describe this process by the $n$-states
wave function $\psi (t)=(b_1(t),\ldots ,b_n(t))$, with $b_i(t)$ is the
amplitude
for the carrier to be in the state $\varphi_j$ on the $i$-impurity site,
which is a solution of
the Schroedinger equation $i\hbar \dot\psi (t)={\cal H}\psi (t)$ for the
initial condition $\psi (0)=(1,0,\ldots ,0)$,
and the effective Hamiltonian is
\begin{equation}
{\cal H}=\left (\matrix{\bar E_0-i\Gamma_1&\Omega_{21}&
\Omega_{31}&\cdots&\Omega_{n1}\cr
\Omega_{12}&\bar E_0&\Omega_{32}&\cdots&\Omega_{n2}\cr
\Omega_{13}&\Omega_{23}&\bar E_0&\cdots&\Omega_{n3}\cr
\cdots&\cdots&\cdots&\cdots&\cdots&\cr
\Omega_{1n}&\Omega_{2n}&\Omega_{3n}&\cdots&\bar E_0-i\Gamma_n\cr}\right )
\label{g1}
\end{equation}
where $\bar E_0=E_0+\epsilon_0$.
Here the coupling with the continuum of the edge states (through the width
$\Gamma$)
is kept only for
the first and for the last impurities, since all the others are farther away
from the edges. The widths $\Gamma_{1,n}$ and
the hopping amplitudes $\Omega_{ij}$ are given by Eq. (\ref{b23}) and
Eq. (\ref{f9}) respectively. We neglected the diagonal energy shift $\Delta_i$
in the Hamiltonian, since $\Delta_i\ll |\Omega|$, Eq. (\ref{ff9}),
although its average value can be included in $\bar E_0$. The amplitude
$\tilde b_n(E)$, which is the Laplace transform of $b_n(t)$, Eq. (\ref{f6}),
defines the probability of the inter-edge transition
\begin{equation}
{\cal R}(E)=4\Gamma_1|\tilde b_n(E)|^2\Gamma_n
\label{g2}
\end{equation}
Notice that the described procedure is very similar to a treatment of resonant
tunneling in multi-well heterostructures\cite{fg}.

It is important to point out that each
impurity generates the quasi-bound state at the same
energy $\bar E_0=E_0+\epsilon_0$,
irrespectively on the impurity position. As a result,
all the diagonal elements in the Hamiltonian (\ref{g1}) are
the same, except for the first and the last one, which
obtain the imaginary part due to the coupling with the edge states.
However, the off-diagonal matrix elements $\Omega_{ij}$ are widely different.
Since $\Omega_{ij}$ drops down very fast for
$|{\mbox{\boldmath $r$}_i}-{\mbox{\boldmath $r$}_j}|>2\ell$, one can expect to
find
only a few paths of a carrier via the $n$-impurities which
would contribute to the inter-edge transport .
Hence, it is not necessary to keep all the amplitudes
$\Omega_{ij}$ in the Hamiltonian (\ref{g1}), but only those which are
associated
with the most probable path of the resonant transitions.

If two or more most probable trajectories
connect the impurities $1$ and $n$ (self-crossing trajectories),
an additional Aharonov-Bohm oscillatory factor would appear in ${\cal R}(E)$.
We now demonstrate on a simple example how our
treatment reproduces this interference effect.
Let us consider four impurities in the channel for a configuration shown
in Fig. 3. We assume that
$|{\mbox{\boldmath $r$}_1}-{\mbox{\boldmath $r$}_2}|=
|{\mbox{\boldmath $r$}_1}-{\mbox{\boldmath $r$}_3}|$ and
$|{\mbox{\boldmath $r$}_2}-{\mbox{\boldmath $r$}_4}|=
|{\mbox{\boldmath $r$}_3}-{\mbox{\boldmath $r$}_4}|$. Therefore $|\Omega_{12}|=
|\Omega_{13}|$ and $|\Omega_{24}|=|\Omega_{34}|$. Since $|\Omega_{14}|\ll
|\Omega_{12}|,|\Omega_{23}|,|\Omega_{24}|$  we neglect the direct transition
between the first and the last impurities by putting  $\Omega_{14}=0$ .
Then the effective Hamiltonian
(\ref{g1}) can be written as
\begin{equation}
{\cal H}=\left (\matrix{\bar E_0-i\Gamma_1&\Omega_{21}&\Omega_{31}&0\cr
\Omega_{12}&\bar E_0&\Omega_{32}&\Omega_{42}\cr
\Omega_{13}&\Omega_{23}&\bar E_0&\Omega_{43}\cr
0&\Omega_{24}&\Omega_{34}&\bar E_0-i\Gamma_4\cr}\right )
\label{g3}
\end{equation}
The probability of the inter-edge transition, ${\cal R}(E)$,
is given by Eq. (\ref{g2}) for $n$=4, where
the 4-state wave function $\tilde\psi (E)=
(\tilde b_1(E),\tilde b_2(E),\tilde b_3(E),\tilde b_4(E))$ is obtained
from the Schroedinger equation in the Laplace-transform form
\begin{equation}
\sum_{j'}(E\delta_{jj'}-{\cal H}_{jj'})\tilde b_{j'}(E)=i\hbar\delta_{1j}
\label{gg3}
\end{equation}
where the r.h.s. reflects the initial condition $\psi (t=0)=(1,0,0,0)$.
By solving Eq. (\ref{gg3}) we find
\begin{equation}
\tilde b_4(E)=i\hbar\frac{(\Omega_{12}\Omega_{23}\Omega_{34}+
\Omega_{13}\Omega_{32}\Omega_{24})-(E-\bar E_0)(\Omega_{12}\Omega_{24}+
\Omega_{13}\Omega_{34})}
{\det [E-{\cal H}]}
\label{g4}
\end{equation}
One finds from Eq. (\ref{g4}) that $\tilde b_4(E)$  has four poles in the
complex
$E$-plane in the vicinity of the Landau level, which generate four resonance
peaks
in the reflection coefficient
${\cal R}(E)=4\Gamma_1|\tilde b_4(E)|^2\Gamma_4$.
The interference effects appears in oscillations of the
numerator in Eq. (\ref{g4}). Indeed,
each term in the numerator corresponds to a certain
path in a carrier inter-edge resonant transition, Fig. 3.
These trajectories do
interfere, since the phases of amplitudes $\Omega_{ij}$ are different.
Using Eq. (\ref{f10}) we obtain
\begin{equation}
|\tilde b_4(E)|=\frac{|\Omega_{12}\Omega_{24}|
\left [( E-\bar E_0)(1+e^{2\pi iBS/\phi_0})-|\Omega_{23}|
(e^{2\pi iBS_1/\phi_0}-e^{2\pi iBS_2/\phi_0})\right ]}{\det [E-{\cal H}]}
\label{gg4}
\end{equation}
where $S\equiv S_{1234}=(x_3-x_2)(y_1-y_4)/2$ is the area enclosed by the two
most
probable trajectories of a carrier, moving from
the impurities 1 to 4, Fig. 3. Correspondingly,
$S_1\equiv S_{123}$ and $S_1\equiv S_{234}$. If $S_1=S_2$
the second term in Eq. (\ref{gg4}) is zero. Then
the reflection coefficient vanishes when the flux $BS=n\phi_0$.

\subsection{Extended states near the Landau levels}

Let us assume that there exists only one most probable chain of non-diagonal
resonant transitions in the Hamiltonian (\ref{g1}). Then neglecting all the
$\Omega_{ij}$ in Eq. (\ref{g1}), which are not associated with this chain
of transitions, one can rewrite the effective Hamiltonian as
\begin{equation}
{\cal H}=\left (\matrix{\bar E_0-i\Gamma_1&\Omega_{21}&
0&\cdots&0&0\cr
\Omega_{12}&\bar E_0&\Omega_{32}&\cdots&0&0\cr
0&\Omega_{23}&\bar E_0&\cdots&0&0\cr
\cdots&\cdots&\cdots&\cdots&\cdots&\cdots\cr
0&0&0&\cdots&\Omega_{n-1,n}&\bar E_0-i\Gamma_n\cr}\right )
\label{g7}
\end{equation}
In this approximation it looks as
the tunneling Hamiltonian for one-dimensional system
with off-diagonal disorder. These systems were studied in many works, see for
instance\cite{day,lif,the,bov,abr}. It was found that the localization length
$L(E)$
diverges in the middle of the band ($E=\bar E_0$), independent
of the probability distribution of off-diagonal disorder\cite{the}. In fact,
this result does not necessarily imply the existence of an extended
state at this energy. It was argued that the wave function would remain
localized
even in the middle of the band, but decaying away from the
localization region only like an exponential of the square root
of the distance\cite{bov,abr}. In our case, however, the situation is
different, since the sequence of non-diagonal
elements in the Hamiltonian (\ref{g7}) is associated with the most probable
path.
Therefore, this sequence is not fully random, but is
a subject of a certain constraint. (Similar
physical situation has been discussed by Lifshitz and Kirpichenkov
in their analysis of tunneling transparency of disordered systems\cite{kirp}).
As a result, the state in the middle of the band may appear
as an extended one. Let us investigate the transport properties
of our system in more detail.

Solving Eq. (\ref{gg3})
with the Hamiltonian (\ref{g7}) we get for $\tilde b_n$
\begin{equation}
|\tilde b_n(E)|=\frac{|\Omega_{12}\Omega_{23}\cdots\Omega_{n-1,n}|}
{|\det [E-{\cal H}]|}
\label{g8}
\end{equation}
The determinant in Eq. (\ref{g8}) can be rewritten in terms of minors and
cofactors as
\begin{equation}
\det [E-{\cal H}]=D_{n}({\cal E})+i\Gamma_1^+
D_{1,n}({\cal E})+i\Gamma_n^-D_{n-1}({\cal E})
-\Gamma_1^+\Gamma_n^-D_{1,n-1}({\cal E})
\label{g10}
\end{equation}
where ${\cal E}=E-\bar E_0$ and $D({\cal E})$ are the minor determinants of
the matrix
\begin{equation}
{\cal D}_{\alpha\beta}={\cal E}\delta_{\alpha ,\beta}+\Omega_{\alpha\beta}
\delta_{\alpha -1,\beta}+\Omega_{\alpha\beta}\delta_{\alpha +1,\beta}
\label{g11}
\end{equation}
so that $D_{n}=\det \{ {\cal D}_{\alpha\beta}\}$ for $\alpha ,\beta=1,\ldots
,n$,
$D_{1,n}=\det \{ {\cal D}_{\alpha\beta}\}$ for $\alpha ,\beta=2,\ldots ,n$,
$D_{n-1}=\det \{ {\cal D}_{\alpha\beta}\}$ for $\alpha ,\beta=1,\ldots ,n-1$,
and
$D_{1,n-1}=\det \{ {\cal D}_{\alpha\beta}\}$ for $\alpha ,\beta=2,\ldots ,n-1$.
Notice that the matrix $\{ {\cal D}\}$ is obtained from the matrix $\{ E-{\cal
H}\}$
by setting $\Gamma_{1,n}=0$.
All the determinants can be evaluated using the recursion relation
\begin{equation}
D_n({\cal E})={\cal E}D_{n-1}({\cal E})-|\Omega_{n-1,n}|^2D_{n-2}({\cal E})
\label{g12}
\end{equation}
where $D_0({\cal E})=1$, and $D_1({\cal E})={\cal E}$.
Finally one obtains for the probability of the inter-edge transition via
$n$-impurities ${\cal R}_n({\cal E})\equiv {\cal R}(E)$, Eq. (\ref{g2})
\begin{equation}
{\cal R}_n({\cal E})
=4\Gamma_1\frac{|\Omega_{12}\Omega_{23}\cdots\Omega_{n-1,n}|^2}
{[D_{n}({\cal E})-\Gamma_1\Gamma_nD_{1,n-1}({\cal E})]^2+
[\Gamma_1D_{1,n}({\cal E})+\Gamma_nD_{n-1}({\cal E})]^2}\Gamma_n
\label{gg2}
\end{equation}
It is clear that the $n$-dependence of ${\cal R}_n({\cal E})$,
determines the localization properties of electron states inside a sample.
For instance, if
these states are localized, ${\cal R}_n({\cal E})$ decreases exponentially
with $n$, and the longitudinal resistance vanishes. We can determine the
localization length $L({\cal E})$ by
\begin{equation}
L^{-1}({\cal E})=-\lim_{n\rightarrow\infty}\frac{1}{n}\ln
\left |{\cal R}_n({\cal E}) \right |
\label{g9}
\end{equation}
One can easily check that this definition of the localization length is
essentially the same as the standard one in terms of eigenstates of the
matrix $\{ E-{\cal H}\}$ with $\Gamma_{1,n}$ in the
Hamiltonian ${\cal H}$, Eq. (\ref{g7}), set to zero\cite{the}.

Consider Eq. (\ref{g12}) for ${\cal E}=0$ (the middle of band).
One finds for $n=2k+1$ that $D_n(0)=D_{n-2}(0)=0$, and
\begin{mathletters}
\label{g13}
\begin{eqnarray}
D_{1,n}(0)&=&(-1)^{\frac{n-1}{2}}|\Omega_{23}|^2|\Omega_{45}|^2\cdots
|\Omega_{n-1,n}|^2
\label{g13a}\\
D_{n-1}(0)&=&(-1)^{\frac{n-1}{2}}|\Omega_{12}|^2|\Omega_{34}|^2\cdots
|\Omega_{n-2,n-1}|^2
\label{g13b}
\end{eqnarray}
\end{mathletters}
Thus using Eq. (\ref{gg2}) we find for the probability of the inter-edge
transition
\begin{equation}
{\cal R}_n(0)=\frac{4\Gamma_1\Gamma_n}{(\Gamma_1\kappa_n^{-1}+
\Gamma_n\kappa_n)^2}
\label{g14}
\end{equation}
where
\begin{equation}
\kappa_n=\left |\frac{\Omega_{12}\Omega_{34}\cdots\Omega_{n-2,n-1}}
{\Omega_{23}\Omega_{45}\cdots\Omega_{n-1,n}}\right |
=\exp [(-{\mbox{\boldmath $r$}}_{12}^2+{\mbox{\boldmath $r$}}_{23}^2-
{\mbox{\boldmath $r$}}_{34}^2+\cdots +{\mbox{\boldmath
$r$}}_{n-1,n}^2)/4\ell^2]
\label{g15}
\end{equation}
and ${\mbox{\boldmath $r$}}_{ij}^2\equiv ({\mbox{\boldmath $r$}}_i-
{\mbox{\boldmath $r$}}_j)^2$. Similarly, for $n=2k$ we obtain
\begin{equation}
{\cal R}_n(0)=\frac{4\Gamma_1\Gamma_n/|\Omega_{n-1,n}|^2}
{\displaystyle\left (\kappa_{n-1}+\frac{\Gamma_1\Gamma_n}{|\Omega_{n-1,n}|^2}
\kappa_{n-1}^{-1}\right )^2}
\label{gg14}
\end{equation}
By applying the central-limit theorem for random distribution of $\Omega_{ij}$
one finds from Eqs. (\ref{g9}), (\ref{g14}) and (\ref{gg14})
that the localization length in the band center diverges,
\begin{equation}
L^{-1}(0)\sim (1/n)|\ln \kappa_n |=\frac{1}{n}\sum_j^n(-1)^j\frac
{{\mbox{\boldmath $r$}}_{j-1,j}^2}{4\ell^2}\sim\frac{1}{\sqrt{n}}
\rightarrow 0
\label{gg15}
\end{equation}
Nevertheless, the electron states in the center of band remain localized, since
${\cal R}_n(0)\sim\exp (-2n^{1/2})\rightarrow 0$ for $n\rightarrow\infty$,
Eqs. (\ref{g14}), (\ref{gg14}).

In our case, however, the distribution of $\Omega_{ij}$ in (\ref{g7})
is not totally random, since the effective Hamiltonian (\ref{g7})
contains only those non-diagonal amplitudes
which generate inter-edge transitions with no appreciable attenuation.
It follows from Eqs. (\ref{g14}), (\ref{gg14})
that $\Omega_{ij}$ have to obey the requirement
$|\kappa_n|\sim 1$, or
$\sum_j(-1)^j{\mbox{\boldmath $r$}}_{j-1,j}^2\sim 4\ell^2$, so that
${\cal R}_n(0)$ would remain of the order of one.
We assume here that the above condition is fulfilled, and as a result the
electron state at ${\cal E}=0$ is an extended one.
Actually, this requirement is much weaker than that
of Lifshitz and Kirpichenkov\cite{kirp} for an appearance of
the "resonance-percolation" trajectories in tunneling through
disordered systems. The reason is that Lifshitz and Kirpichenkov looked for
a band of extended states, and therefore they required
approximately the same distance between impurities in the chain.
In a contrast, our condition is a sufficient one only for a single
extended state, where all other electron states are localized.

\subsection{Localization properties of electron states for ${\cal E}\not =0$}

In order to investigate the electron transport along the chain
of randomly distributed impurities, restricted by
the constraint $\kappa_n\sim 1$, it is easier to start with the opposite
case, where all impurities are aligned and therefore
the non-diagonal amplitudes in the Hamiltonian (\ref{g7}) are the same,
$\Omega_{ij}=\Omega$. Then it is rather obvious that
the conducting band appears around ${\cal E} =0$. It can be seen immediately
from
Eq. (\ref{g12}) that in this case
\begin{equation}
D_n({\cal E})=\Omega^n\frac{\sin [(n+1)\alpha ]}{\sin (\alpha )}
\label{g16}
\end{equation}
where $\cos (\alpha ) ={\cal E}/2\Omega$. Substituting Eq. (\ref{g16}) into
Eq. (\ref{gg2}) and taking for the simplicity $\Gamma_1=\Gamma_n=\Omega$ we
obtain for probability of the inter-edge transition across a sample
\begin{equation}
{\cal R}_n({\cal E})=\frac{4\Omega^2-{\cal E}^2}{4\Omega^2-{\cal E}^2
\cos^2(n\alpha )}
\label{g17}
\end{equation}
Therefore all the states for $|{\cal E}|\leq |\Omega |$ are indeed extended.

Now let us displace the impurity '$k$' from its position so that
$\Omega_{k-1,k}\not =\Omega$ and $\Omega_{k,k+1}\not =\Omega$. We assume
for the definiteness that $|\Omega_{k-1,k}|,|\Omega_{k,k+1}|\ll |\Omega |$.
As a result the impurity '$k$' would generate
the reflection of the Bloch waves inside the band.
For the energy ${\cal E}\ll \Omega$, one finds that the situation
resembles resonant tunneling through a double-barrier structure, namely
the Bloch waves inside the band can pass the impurity '$k$' only
via the virtual state at ${\cal E}=0$.
Then the probability of inter-edge transition can be written as
\begin{equation}
{\cal R}_n^{(k)}({\cal E})=\frac{4\gamma_{k-1}\gamma_{k+1}}{{\cal E}^2+
(\gamma_{k-1}+\gamma_{k+1})^2}
\label{g18}
\end{equation}
(c.f with Eq. (\ref{b25})), where
\begin{mathletters}
\label{g19}
\begin{eqnarray}
\gamma_{k-1}=2\pi\varrho |\Omega_{k-1,k}|^2=\frac{|\Omega_{k-1,k}|^2}{|\Omega
|}
\label{g19a}\\
\gamma_{k+1}=2\pi\varrho |\Omega_{k,k+1}|^2=\frac{|\Omega_{k,k+1}|^2}{|\Omega
|}
\label{g19b}
\end{eqnarray}
\end{mathletters}
Here $\varrho =(2\pi |\Omega |)^{-1}$ is the density of the band states.

The above result clarifies the meaning of the constraint $\kappa_n\sim 1$
which is the necessary one
for the appearance of an extended state at ${\cal E}=0$. Indeed, the
maximal value of the
transition probability is reached at the resonance energy ${\cal E}=0$ and
only for $\Omega_{k-1,k}/\Omega_{k,k+1}=1$, as always happened
in the case of resonant scattering.
Since all the other $\Omega_{ij}$ are the same, the above condition is
equivalent to $\kappa_n = 1$. The
displacement of an another impurity would create the second
double barrier structure along the chain. By approximating
the total transmission probability as a product\cite{the,bush}
${\cal R}_n^{(k_1)}{\cal R}_n^{(k_2)}$,
one finds that also in this case the maximal probability is reached for
$\Omega_{k_1-1,k_1}/\Omega_{k_1,k_1+1}=1$ and
$\Omega_{k_2-1,k_2}/\Omega_{k_2,k_2+1}=1$, so that
$\kappa_n$ remains to be one. By continue this procedure of increasing disorder
one finds that the resonance at ${\cal E}=0$ by itself
would not provide the absence of an attenuation along the chain. One
needs also to keep $\Omega_{k-1,k}/\Omega_{k,k+1}=1$, in an accordance with
the requirement of $\kappa_n\sim 1$.

Now let us estimate the transmission probability for ${\cal E}\not =0$,
by imposing  $\Omega_{k-1,k}=\Omega_{k,k+1}$, so that the state ${\cal E}=0$
is the extended one. (It is quite clear from the above result that if the state
${\cal E}=0$ is localized, all the other states for ${\cal E}\not =0$
are localized too). Using Eqs. (\ref{g18}),(\ref{g19})
we obtain
\begin{equation}
{\cal R}_n({\cal E})=\prod_k\frac{1}{\displaystyle{{\cal E}^2\frac{|\Omega |^2}
{4|\Omega_k|^4}+1}}
\label{g20}
\end{equation}
where $\Omega_{k-1,k}=\Omega_{k,k+1}\equiv\Omega_k$,
Therefore the electron states for ${\cal E}\not =0$ are localized with the
localization length, Eq. (\ref{g9})
\begin{equation}
L^{-1}=C\ln\left ( 1+\frac{{\cal E}^2}{4\langle\Omega\rangle^2}\right )\sim
\frac{{\cal E}^2}{\langle\Omega\rangle^2}
\label{g21}
\end{equation}
and the transmission probability across a sample decreases exponentially with
$n$, i.e.  ${\cal R}_n({\cal E})\sim
\exp [-({\cal E}^2/\langle\Omega\rangle^2)n]$.

Although we concentrated on electron transport
inside the Landau mode $N=0$, it is clear that the results remain
the same for any of Landau modes. Our analysis of the
localization properties was based only on the off-diagonal disorder in
the effective Hamiltonian (\ref{g7}) and on the constraint $\kappa_n\sim 1$,
Eq. (\ref{g15}). Therefore the exponential is universal and
the localization length near the Landau level
$N$ band center can be written as
\begin{equation}
L_N^{-1}(E)\sim (E-\bar E_N)^2
\label{g22}
\end{equation}
where $\bar E_N=E_N+\epsilon_N$ and  $\epsilon_N$ is the energy of the
resonance near the Landau level $E_N$. Thus the critical energy $\bar E_N$
is shifted away from the exact center of the
Landau level $E_N$ (as also follows from the numerical
calculations\cite{liu,aoki}). Its position depends on the type of impurities
in a sample: for attractive impurities it is below the Landau level, and
above the Landau level for repulsive impurities.

We also like to point out that the peak value of the
reflection coefficient ${\cal R}_{max}={\cal R}(0)$,
Eqs. (\ref{g14}),(\ref{gg14}),
averaged over all possible resonant-percolation trajectories,
determines the maximum of the total (two-terminal)
resistance ${\cal G}^{-1}$,
Eq. (\ref{a11}). Since our derivation of Eqs. (\ref{g14}),(\ref{gg14})
does not depend on a particular Landau mode $N$, we expect that the peak value
of two-terminal resistance remains the same for all Landau levels.

The particular behavior of the electron states near the
the Landau level center, where one (resonance) level is extended and others
are localized, is a consequence of the off-diagonal disorder of the
effective Hamiltonian (\ref{g7}). The diagonal energy shift, due to
resonance transitions between different impurities would violate the diagonal
order. However, in a case of short range impurities
the diagonal shift $\Delta_j$, Eq. (\ref{ff9}), is small and can be
can be neglected. When the potential range increases the renormalization
of the resonance levels may be appreciable. Nevertheless, the diagonal disorder
is determined by $\Delta_j-\langle\Delta\rangle$, where $\langle\Delta\rangle$,
is the mean value of the diagonal energy shift, while the off-diagonal
disorder,
is determined by the overlap of the quasi-bound states near impurity sites,
$\Omega_{ij}$. Since in strong magnetic field
these states are concentrated near the impurity centers even for long range
potentials, Eq. (\ref{ee7}), one finds that $\Omega_{ij}$ would fluctuate
much strongly than $\Delta_j-\langle\Delta\rangle$. In this case we may
expect that the off-diagonal disorder would still play a dominant role in
a determination of localization properties of electron states.

Finally we would like to mention that in the case of more than one
resonant-percolation trajectories connect the same initial and final
impurity sites (as in Eq.(\ref{g3})) (self-crossing trajectories),
one cannot re-arrange the effective
Hamiltonian in a form of the off-diagonal nearest-neighbor
coupling. In this case the Aharonov-Bohm interference-effects may play an
essential role and the value of the exponential in Eq. (\ref{g22})
can be modified. This effect needs a special investigation
and is not a subject of this paper.

\section{Summary}

In this paper we presented a systematic study of the resonant scattering
on impurities of 2D electron gas in the presence of a strong magnetic
field. It follows from our analysis that this process can play a dominant role
in delocalization of the electron states near the Landau levels.

Starting with a detailed analysis of the resonant scattering on
a single impurity we demonstrated that any
no matter how weak impurity generates resonance states in the
vicinity of the Landau levels: an attractive impurity produces the resonances
below
the Landau levels, and a repulsive impurity -- above the Landau levels. Our
method
can be applied to any type of impurities. In particular, we found
simple analytical expressions for the resonance energy and the partial
widths for any short range impurity and for
long range impurity of a Gaussian type. Although these expressions look
rather similar for attractive and repulsive impurities, the
mechanism of the inter-edge transition
is quite different: The attractive impurity gives rise to the inter-edge
transitions via coupling with the resonant state in the nearest non-propagating
Landau mode. In contrast, the inter-edge transition generated by
a repulsive impurity takes place in the same
propagating Landau mode. As a result, the transition probability via
a repulsive impurity is in general larger than that produced by an
attractive impurity. For instance, in the case of an attractive impurity the
value of the resonance peak in the longitudinal resistance
drops down very rapidly, when the impurity is away from the center of
a sample. On the other hand, the value of the resonance peak generated by
a repulsive impurity is less sensitive to the impurity position.

In the case of wide samples the inter-edge transport
is generated by a multiple scattering of carriers on different
impurities. It is known that in the absence of magnetic field
the electron states of disordered 2D system are localized. Therefore the
question is of how a strong magnetic field can delocalize the
states, so that the inter-edge current could flow across a sample.
Our analysis of the resonant scattering on impurities suggests the following
reasons. First, each impurity in a strong magnetic field
generates the resonance at the same energy which
depends only on the type of impurity. The imaginary part of the resonance
energy (the width) is appreciable only
for those impurities which are close to the sample edges, and it is
negligible for impurities in the bulk.
Second, the quasi-bound states are strongly
localized near the impurity center (even for a long range impurity, as follows
from our analysis of the Gaussian-type impurity). As a result, the diagonal
energy shift plays no essential role in the transport properties of carriers,
in
comparison with the off-diagonal transition amplitudes.
(In a case of short range impurities the diagonal energy
shift can be neglected at all).
Therefore the system resembles eventually the one with off-diagonal disorder.

The analysis of 1D system
with the off-diagonal disorder shows that the localization properties
of the level in the center of the band (the resonance energy)  differ from that
of the other states. One finds that the wave function of this state is
localized,
but it decays
only like an exponent of the square root of the distance. It is enough to
impose a weak constraint on the position of impurities in the chain in order to
delocalize this state. In our case of
2D disordered system, the transport takes place along the most probable
trajectories
(the resonant percolation trajectories). If we neglect the contributions
from self-crossing trajectories, the problem is reduced
to that of the 1D system  with off-diagonal
disorder. The requirement for the resonance-percolation trajectory
in 2D system delocalizes the level in the center of the band.
We found this loose constraint on the
impurity positions in the chain, and demonstrated
that all other levels except for the central one, remain localized.
The analysis of
the localization properties can be done analytically. We estimated the
behavior of the localization length near the band center under
the above constraint
on the impurity positions, and obtain the square power-like divergence of the
localization. The latter would remain the same for any of Landau levels,
as well as the peak value of the {\em two-terminal} resistance.

It is clear that this analysis can give only an estimation of the critical
exponent.
For instance, the neglected contribution of the self-crossing
resonant-percolation trajectories may change the value of this exponent.
These and other effects are not considered here and need further investigation.

\section{Discussion}

It is known that all the states are localized for a 2D electronic
system with arbitrary amount of disorder. However, when a uniform magnetic
field is applied perpendicular to the system, there is an indication that
extended electron states exist at only one energy within the broadened
Landau level. Other states should be localized. Despite various attempts the
nature of the delocalization is not clear yet. In Section 5 we
proposed a possible mechanism of the delocalization and predicted universal
power-law divergence of the localization length.
Therefore it is interesting to compare our results with
the existing approaches.

Two different theories exist which predict the universal power-low
divergence $(E-\bar E_N)^{-\nu}$ of the localization length
for noninteracting electrons in the presence of
a high magnetic field. The first based on a classical percolation theory,
considers the extreme high-magnetic-field limit
when the magnetic length is much less than the correlation length of
the disorder potential\cite{trug}. This approach predicts the value of
the critical exponent $\nu =4/3$.
The second theory is based on an approximate
reduction of the problem to nonlinear $\sigma$-model\cite{pruis}.
This model is applicable when the magnetic length is much
larger than the correlation length of the disorder potential. The value
of the critical exponent, however, is not known for the non-linear
$\sigma$-model.

Existing numerical calculations also establish the power-low
divergence of the localization length. The value of $\nu\approx 2.2-2.3$
was obtained for the lowest Landau level in the case of short-range
white-noise random potential\cite{huck} and for $\delta$-function or
Gaussian-type disorder\cite{liu}. Still, the numerical analysis does not
illuminate the origin of the localization mechanism.

Our analytical study of this problem given for randomly distributed
{\em identical} impurities, also demonstrates
the appearance of a delocalized state and the universal
power-low divergence of the localization length with $\nu\approx 2$.
An important advantage of our approach is that it clearly displays the
delocalization mechanism.
Paradoxically, the extended state appears as a result of the
localization of an electron near impurity sites due to
strong magnetic field. Since the energy of these localized
states is the same for all impurities, the disorder takes place
only in non-diagonal matrix elements (2D off-diagonal disorder).
Then  as explained in Section 5, the resonant percolation trajectories appear
which lead to a delocalization of one particular level. In fact, one
can prove the existence of these trajectories even for an arbitrary small
density of impurities, providing the size of a sample is large enough.
The proof will be given in a separate publication as well as
an extension for a general case of random disorder (non identical impurities).

Since we considered here the case of identical impurities, the position
of the critical energy $\bar E_N$ is slightly shifted from the center
of the corresponding Landau level, $E_N$ (c.f. with finding
in\cite{liu,aoki}). It lies in an agreement with the result
of\cite{pruis}, which predicts only the universality of the exponent,
but not the critical energy. The latter can be slightly changed, depending on
properties of a sample.

The scaling behavior of the localization length is closely related to the
remarkable temperature dependence of the half-width of peaks in
the longitudinal resistance. The experiments of Wei {\em et. al.}\cite{wei}
showed that this quantity vanishes as $\sim T^{\alpha}$,
where $\alpha\approx 0.42$ is the same for the integer and for the
fractional QHE. Despite the fact that
$\alpha$ is believed to involve the inelastic scattering length,
which is not understood in the high-field regime, the universality of
$\alpha$ indicates on the same localization mechanism for the integer and
the fractional QHE. Indeed it was explicitly established by using the
composite fermions\cite{jain1,jain2} or the anyons\cite{schm} approaches
to the fractional QHE. It is rather clear that in this framework
our approach would also predict the same localization mechanism for the integer
and the fractional QHE, except for half-filled Landau level.
In the latter case the system would be equivalent to the
spinless fermions in {\em zero} magnetic field\cite{halp}.
Hence, it cannot be effectively reduced to one described in
terms of off-diagonal 2D disorder, and therefore our analysis
would not be more valid.

\section{Acknowledgments}

I owe special thanks to A. Aronov for useful discussions and important
suggestions. I am also grateful to  A. Finkel'stein, Y. Gefen, S. Iordansky,
Y. Levinson and A. Kamenev for useful discussions.

\appendix

\section{square well confined potential in the magnetic field}

Consider Eq. (\ref{a3})
for the Landau-mode wave functions in the confining potential:
$W(y)=0$ for $|y|\le a$ and $W(y)=+\infty$ for $|y|>a$.
The general solution of
this equation can be written as a linear combination
of parabolic cylinder functions, $D_{\nu}(z)$ and $D_{\nu}(-z)$,
\begin{equation}
\Phi_{nk}(y)=C_+D_{\nu}(z)+C_-D_{\nu}(-z)
\label{ap1}
\end{equation}
where $\hbar\omega_c (\nu +1/2)=E_{nk}$ and $z=\sqrt{2} (y-
\ell^2k)/\ell$.
\begin{eqnarray}\label{ap2}
D_{\nu}(z)=\frac{2^{\nu /2}e^{-z^2/4}}{\sqrt{\pi}}
& &\left [\Gamma\left (\frac{1+\nu}{2}\right )\cos (\pi\nu /2){_1F_1}
\left (-\frac{\nu}{2};\frac{1}{2};\frac{z^2}{2}\right )\right.\nonumber\\
& &\left.+\sqrt{2}z
\Gamma\left (1+\frac{\nu}{2}\right )\sin (\pi\nu /2)
{_1F_1}\left (\frac{1-\nu}{2};\frac{3}{2};\frac{z^2}{2}\right )\right ]
\end{eqnarray}
Here $_1F_1$ is the confluent hypergeometric (or Kummer) function.
For large positive $z\gg 1$ and $z\gg |\nu|$
\begin{equation}
D_{\nu}(z)\sim e^{-z^2/4}z^{\nu}
\left [1 -\frac{\nu (\nu -1)}{2z^2}\pm \cdots\right ]
\label{ap3}
\end{equation}
and if $z$ is large and negative $z\ll -1$ and $z\ll -|\nu |$
\begin{eqnarray}\label{ap4}
D_{\nu}(z)\sim e^{-z^2/4}z^{\nu}
& &\left [1-\frac{\nu (\nu -1)}{2z^2}\pm \cdots\right ]\nonumber\\
-& & \frac{\sqrt{2\pi}}{\Gamma (-\nu )}e^{\nu\pi i}e^{z^2/4}z^{-\nu -1}
\left[1+\frac{(\nu +1)(\nu +2)}{2z^2}\pm \cdots\right ]
\end{eqnarray}
By supplementing the boundary conditions, $\Phi_{nk}(\pm a)=0$
one obtains the energy spectrum from the equation
\begin{equation}
D_{\nu}(z_+)D_{\nu}(z_-)-D_{\nu}(-z_+)D_{\nu}(-z_-)=0
\label{ap5}
\end{equation}
where $z_{\pm}=(a/\ell )\pm k\ell$. Let us consider the region of
$|z_{\pm}|\gg 1$, where one can use the asymptotic representation
of the parabolic cylinder functions. Substituting
Eqs. (\ref{ap3}),(\ref{ap4}) into Eq. (\ref{ap5}), and keeping only the
leading terms of the expansion, one obtains
\begin{equation}
z_+^{2\nu +1}e^{-z_+^2/2}+z_-^{2\nu +1}e^{-z_-^2/2}=\sqrt{\frac{2}{\pi}}
\frac{1-e^{-2\pi i\nu}}{2i}\Gamma (1+\nu )
\label{ap6}
\end{equation}
Taking $\nu =n+\alpha_n$ and expanding
Eq. (\ref{ap6}) in power of $\alpha_n$
we get
\begin{equation}
\alpha_n=\frac{2^n}{\sqrt{\pi} n!}\left [\left (\frac{a}{\ell}-k\ell
\right )^{2n+1}\exp\left (-\frac{(a-k\ell^2)^2}{\ell^2}\right )
+\left (\frac{a}{\ell}+k\ell
\right )^{2n+1}\exp\left (-\frac{(a+k\ell^2)^2}{\ell^2}\right )\right ]
\label{ap7}
\end{equation}
Using ${\cal K}_n(k)
=\hbar\omega_c\alpha_n$, we arrive to
Eq. (\ref{aa1}) for the energy spectrum in the confined square-well
potential for
$|a\pm k\ell^2|\gg \ell$. One also finds that
the eigenfunctions, Eq. (\ref{ap1}), are well approximated by
by the harmonic oscillator wave functions, Eq. (\ref{a5}), in the
region $|a\pm k\ell^2|\gg \ell$.

\section{hopping transitions between two impurities}

Consider two impurities, each of them generates a bound state
with the energy $E_1$ and $E_2$ respectively
\begin{mathletters}
\label{bp1}
\begin{eqnarray}
({\cal K}+V_1)\psi_1=E_1\psi_1
\label{bp1a}\\
({\cal K}+V_2)\psi_2=E_2\psi_2
\label{bp1b}
\end{eqnarray}
\end{mathletters}
Here ${\cal K}$ denotes the kinetic energy term and
$V_{1,2}$ are the impurity potentials. We assume that
$E_1\sim E_2$. These energies are obviously not the eigenvalues
of the whole Hamiltonian, when the two potentials are
"switched on". The exact energies are obtained from the
the Schroedinger equation
\begin{equation}
({\cal K}+V_1+V_2)\psi =E\psi
\label{bp2}
\end{equation}
Let us rewrite this equation in the integral form as
\begin{equation}
\psi =\psi_1+\tilde G_1V_2\psi =\psi_1 +\tilde G V_2\psi_1
\label{bp3}
\end{equation}
where the Green's functions
\begin{equation}
\tilde G_1=\frac {Q_1}{E-{\cal K}-V_1}
\label{bp4}
\end{equation}
and
\begin{equation}
\tilde G=\tilde G_1+\tilde G_1V_2\tilde G
\label{bp5}
\end{equation}
Here $Q_1$ is the projection operator which excludes the state
$|\psi_1><\psi_1|$ from the spectral representation of
$G_1=(E-{\cal K}-V_1)^{-1}$.
The bound state energy $E$ is given by
\begin{equation}
E=E_1+<\psi_1|V_2|\psi_1>+<\psi_1|V_2\tilde GV_2|\psi_1>
\label{bp6}
\end{equation}

Let us assume that the impurities are separated far enough so that the
wave functions $\psi_{1,2}$ are small in the regions of the potentials
$V_{2,1}$, respectively. Then one can approximate
$Q_1\cong 1$ in the region of the potential $V_2$, and therefore
$V_2\tilde GV_2\cong
V_2GV_2$, where
\begin{equation}
G=\frac{1}{E-{\cal K}-V_1-V_2}=G_2+G_2V_1G_2+G_2V_1G_2V_1G_2+\cdots
\label{bp7}
\end{equation}
is the total Green's function, and
$G_2=(E-{\cal K}-V_2)^{-1}$. Using the spectral representation
of the Green's function $G_2$ and Eq. (\ref{bp7})
we get for $G_2$ and $G$ in the energy region
$E\sim E_{1,2}$
\begin{mathletters}
\label{bp8}
\begin{equation}
G_2\rightarrow\frac{|\psi_2><\psi_2|}{E-E_2}
\label{bp8a}
\end{equation}
\begin{equation}
G\rightarrow\frac{|\psi_2><\psi_2|}{E-E_2-<\psi_2|V_1|\psi_2>}
\label{bp8b}
\end{equation}
\end{mathletters}
Substituting this result into Eq. (\ref{bp6}) we get
\begin{equation}
E=E_1+<\psi_1|V_2|\psi_1>+\frac{<\psi_1|V_2|\psi_2><\psi_2|V_2|\psi_1>}
{E-E_2-<\psi_2|V_1|\psi_2>}
\label{bp9}
\end{equation}
This equation can be identically rewritten as
\begin{equation}
\det\left (\matrix{E-E_1-\Delta_1&\Omega\cr
\Omega&E-E_2-\Delta_2\cr}\right )=0
\label{bp10}
\end{equation}
where $\Delta_1=<\psi_1|V_2|\psi_1>$ and
$\Delta_2=<\psi_2|V_1|\psi_2>$ are the diagonal energy shifts,
and $\Omega =<\psi_1|V_2|\psi_2>$ is the hopping transition amplitude
between two impurities. Using the Schroedinger equation  one obtains
useful formula for the hopping amplitude
\begin{equation}
\Omega =
<\psi_1|V_2|\psi_2>\cong <\psi_1|V_1|\psi_2>\cong
<\psi_1|\bar E-{\cal K}|\psi_2>
\label{bp11}
\end{equation}
where $\bar E=(E_1+E_2)/2$. Notice that $\Omega\gg\Delta_{1,2}$.

\begin{figure}
\caption{(a)  Energy spectrum for 2D electron gas in a strong magnetic
field for a rectangular confining potential. (b) Effective 'kinetic
energy' for different Landau bands.}
\end{figure}
\begin{figure}
\caption{Schematical illustration of the classical hopping motion along
the boundaries of a sample and the inter-edge transitions via different
impurities.}
\end{figure}
\begin{figure}
\caption{Different optimal trajectories for the inter-edge transition
via four impurities.}
\end{figure}

\begin{references}
\bibitem{com1} We are not entering into the question of how to obtain the
longitudinal and the Hall resistance from ${\cal G}$. This problem has been
extensively discussed in the literature.  For instance, one
can use the B\"uttiker formalism\cite{but}.
\end{references}
\end{document}